# Local Structure of Multiferroic $TbMn_2O_5$: Evidence for an Anomalous Tb –O Distribution

T. A. Tyson[1,3], M. Deleon[1], S. Yoong[2,3], and S.-W. Cheong[2,3]
[1]Department of Physics, New Jersey Institute of Technology, Newark, NJ 07102
[2]Department of Physics and Astronomy, Rutgers University, Piscataway, NJ 08854
[3]Rutgers Center for Emergent Materials, Rutgers University, Piscataway, NJ 08854

## Abstract

The temperature dependent local structure of $TbMn_2O_5$ was determined by x-ray absorption spectroscopy. An anomalous Tb-O distribution is found. At high temperature it is broad but resolves into two distinct peaks below ~180 K. The distributions sharpen below the Tb magnetic ordering temperature (~10 K). The distortions in the Tb-O distribution, away from the Pbam structure, are consistent with rotations of the $MnO_x$ polyhedra about the c-axis and suggest that Tb-O bond polarization may play a significant role in the observed ferroelectric properties of this system.

# I. Introduction

Magnetic and ferroelectric materials have been extensively studied for several decades from both applied and basic research perspectives. This work has resulted in the development of devices ranging from transformers to hard-drive read–head sensors and data storage media [1,2,3]. These materials are of deep interest from a fundamental science perspective. They reveal an intimate coupling of electron, spin and lattice degrees of freedom.

For the classic ferroelectric materials, the simple perovskite $ABO_3$ systems such as ($PbTiO_3$) have been extensively studied [2]. These materials possess a spontaneous net electric polarization (**P**) below the so-called ferroelectric transition temperature and exhibit hysteresis in the presence of an externally applied electric field in analogy to that found in magnetic systems. From a microscopic perspective these materials fall into two main classes. One class (displacive) has a nonzero |**P**| due to off-center atomic displacements. Alternatively, order-disorder transitions have been proposed to explain the finite |**P**| at low temperatures.

In the ideal displacive ferroelectric system, symmetry reduction occurs on entering the ferroelectric phase from the high temperature paraelectric phase. In $PbTiO_3$, atomic off-center displacements of Ti and Pb are ~0.2 Å, with onset 190 K above $T_c$ [4]. For the order-disorder system [5] the B site atom is thought to randomly occupying each of eight equivalent off-center sites above the transition with a preferred occupancy occurring below the transition temperature. Detailed analysis by microscopic probes such as x-ray absorption spectroscopy and nuclear magnetic resonance measurements are leading to a more complex picture in which both types of transitions are seen to exist in these systems with dynamic hoping of atoms to equivalent off-center positions being turned on at higher temperatures allowing the atoms to overcome the barriers [6]. In general both disorder and displace character must be considered in a given system (see for example Ref. [7]).

Magneto-electric multiferroics are a class of materials which are simultaneously ferroelectric and ferromagnetic [3,8]. The possibility of coupling of the magnetic and electric properties may enable new functions such as the ability to store data as both magnetic and electrical bits and the ability to write ferroelectric bits with magnetic fields. Not many multiferroelectric systems have been observed. It has been argued by Hill [3] that while the 3d occupancy on the B site in $ABO_3$ systems creates unpaired electrons needed for magnetism it also stabilizes inversion center preserving Jahn-Teller distortions. Magnetoelectric effects have also been explored in systems such as $Ti_2O_3$, $GaFeO_3$, boracite, $TbPO_4$, $BiFeO_3$ and $BiMnO_3$ [8]. However, the magneto-electric coupling in these systems is weak and possibly originats from mechanisms which differ from the classic ferroelectric systems

Recently, Kimura *et al.* [9] discovered a large magnetoelectric and magnetocapacitance effects induced by magnetic fields in the simple perovskite system $TbMnO_3$. The application of a magnetic field along the b-axis suppresses **P** along the b-axis but enhances it along the a–axis. The low temperature (below 40 K) spontaneous |**P**| approaches ~ 80nC/cm$^2$ (~10$^2$ times lower than room temperature $BaTiO_3$).

More recently, Hur *et al.* [10] discovered reversible low temperature (3 K) switching in $TbMn_2O_5$. By sweeping the magnetic field from 0 to 2 T, |**P**| passes through

zero and attains a value with magnitude ~equal to the zero field value (~40nC/cm$^2$). Reducing the field to zero recovers the initial state.

The low temperature behavior of his system is quite complex [11]. The Mn sites order antiferromagnetically at 43 K followed by the onset of ferroelectricity at 38 K. At 33 K (near maximum |**P**|) the magnetic propagation vector locks into the commensurate value (1/2,0,1/4) then becomes incommensurate on cooling below 24 K. A large enhancement in ordering of the Tb moments occurs at 10 K.

Measurement of the thermal expansivity ($\alpha$) along the a, b and c axes reveals jumps at the magnetic and electric ordering temperatures [12]. The change in crystal dimension by this bulk measurement is of order ($\Delta L/L$) $10^{-6}$. Anomalous changes have also been found in the atomic displacement factors of the O(2), $Mn^{3+}$, and Tb sites [11]. Interestingly, high-resolution synchrotron x-ray diffraction measurements of high-order reflections in $YMn_2O_5$ reveal no change in lattice parameters above the level of ~$10^{-4}$ Å [13].

The long-range room temperature structure (Pbam) of the $ReMn_2O_5$ system (where Re is a rare earth atom) was systematically explored by Alonso *et al* and Kagomiya *et al*. [14]. The $Mn^{4+}O_6$ octahedra (Mn1 sites) form infinite chains parallel to the c-axis of the orthorhombic Pbam cell. These chains are cross-linked by $Mn^{4+}O_5$ pyramids (Mn2 sites) which form edge-sharing dimmers ($Mn^{4+}{}_2O_8$). The Re atoms are eight-fold coordinated to oxygen atoms. We note that the inversion center in this Pbam space group is inconsistent with a finite polarization.

In order to understand the mechanism responsible for the finite polarization from a microscopic perspective, we have examined the temperature dependent changes in the local structure of $TbMn_2O_5$ between 3 K and 470 K by x-ray absorption spectroscopy. An anomalous Tb-O distribution is found. At high temperature it is broad but resolves into two distinct peaks below ~180 K. The distribution sharpens below the Tb ion magnetic ordering temperature (10 K). The distortion of the Tb-O distribution is inconsistent with the reported crystal structure suggesting a lower symmetry space group and possibly a larger unit cell.

## II. Experimental Methods

Polycrystalline samples of $TbMn_2O_5$ were prepared by solid state reaction in oxygen. X-ray absorption samples were prepared by grinding and sieving the material (500 mesh) and brushing it onto Kapton tape. Layers of tape were stacked to produce a uniform sample for transmission measurements with jump $\mu t \sim 1$. Spectra were measured at the NSLS beamline X19A at Brookhaven National Laboratory. Measurements were made on warming from 3 K in He vapor in the cryostat of a superconducting magnet [15]. Two to six scans were taken at each temperature. The uncertainty in temperature is < 0.1 K. A Mn foil reference was employed for energy calibration. The reduction of the x-ray absorption fine-structure (XAFS) data was performed using standard procedures [16]. The energy $E_0$ (k=0) was chosen to be 6555.3 eV and 7518.7 eV, for the Mn K and Tb L3 spectra, respectively. Representative XAFS data at 300 K are shown in Fig. 1 for both the Mn K-edge (thin line) and the Tb L3-Edge (thick line).

To treat the atomic distribution functions on equal footing at all temperatures the spectra were modeled in R-space by optimizing the integral of the product of the radial distribution functions and theoretical spectra with respect to the measured spectra.

Specifically the experimental spectrum is modeled by, $\chi(k) = \int \chi_{th}(k,r)\, 4\pi r^2 g(r) dr$ where $\chi_{th}$ is the theoretical spectrum and g(r) is the real space radial distribution function based on a sum of Gaussian functions ($\chi(k)$ is measured spectrum) [17] at each temperature. Theoretical spectra for atomic shells [18] were derived from the Pbam crystal structure [14]. For the Mn K-Edge, fits of the Mn-O distribution were confined to the k-range $2.64 < k < 12.15$ Å$^{-1}$ and to r-range $0.89 < R < 2.04$ Å while for the Tb L3-Edge both the Tb-O and Tb-Mn distributions were modeled: k-range $2.642 < k < 12.107$ Å$^{-1}$ and r-range $1.13 < R < 3.38$ Å . For the Mn K-edge, the coordination number for the average Mn-O shell was fixed at 5.5 (average of $Mn^{3+}$ and $Mn^{4+}$ site) while varying the width ($\sigma^2$) and positions (R) of the Gaussian component of the radial distribution functions. For the Tb L3-Edge both to Tb-O first shell and the Tb-Mn second shell were fit. For the first shell the total coordination was held fixed at the crystallographic value of N=8 for one-shell fits (above ~180 K) and $N_{Short} + N_{Long} = 8$ for two-shell fits below ~180 K. The Gaussian widths and positions were fit for each component. For the Tb-Mn shell the coordination number was fixed at 6 while $\sigma^2$ and R were fit. Representative fits are shown in Fig. 2 and extracted parameters for the more complex Tb-O distribution are shown in Fig 5. Errors reported are based on the covraience matrices of the fits (stability of fits) and are upper limits to the errors obtained from the experimental spread in consecutively measured data scans. The temperature dependence of the Mn-O and Tb-Mn Debye-Waller factors ($\sigma^2$) was modeled by an static contribution ($\sigma_0^2$) plus a single parameter ($\theta_E$) Einstein model using the functional form $\sigma^2(T) = \sigma_0^2 + \frac{\hbar}{2\mu k_B \theta_E} \coth(\frac{\theta_E}{2T})$ [19], where μ is the reduced mass for the bond pair. This simple model represents the bond vibrations as harmonic oscillations of a single effective frequency proportional to $\theta_E$. It provides an approach to characterize the relative stiffness of the bonds.

## III. Results and Discussion

We now examine the structure about the Mn and Tb sites, starting with the Mn-O distribution about the average Mn site. In Figs. 3 we display the extracted temperature dependence of $\sigma^2$ for the distribution of Mn-O distances and a fit to the Einstein model. The total distribution contains five distinct narrowly spaced shells (according to the Pbam model) and is well fit by a narrow Gaussian distribution yielding a low temperature Debye-Waller factor (static ($\sigma_0^2$ )+ temperature dependent) of ~ 0.003 Å$^2$. In addition, to the conforming to the diffraction derived structure with respect to the Mn site, the main observation from these fits, which models the average Mn-O correlations in both the $Mn^{4+}O_6$ ocatahedra and $Mn^{3+}O_5$ pyramids, is that these polyhedra are rigid as can be seen from the large Einstein temperature (~750 K). No anomalous changes in $\sigma^2$ or the average distance (no greater than 0.01 Å) were found over the range of temperatures studied including the region below 40 K where magnetic ordering and ferroelectricity are observed.

With respect to the Tb site the picture is quite different. In Fig. 4(a), we show the Fourier transform of the Tb L3 spectra taken at 20 K (thin line) and 300 K (thick line). These spectra are compared with a model spectrum, based on the Pbam structure [14] including a reasonable global Debye-Waller factor (same for all shells) of $\sigma^2$ =0.0023 Å$^2$ for low temperature (dotted line). The Tb-Mn and higher shells (Tb-Tb and Tb-Mn shells) are matched by this simple model. However, note that the Tb-O shell exhibits a significant deviation from the diffraction derived model. The low-temperature (thin line) data reveal a broad, suppressed and asymmetric function compared to the model - suggesting a complex distribution.

While the Tb-Mn distribution is well modeled by a single Gaussian distribution yielding a Einstein temperature of ~300 K (Fig. 4(b)), the first shell about Tb admits no such simple solution. The Tb-O distribution can not be modeled by the narrow distribution predicted by the crystallographic structure at any of the temperatures recorded in our measurements set. In fact, we find the rough value $\sigma^2$ ~ 0.0078 Å$^2$ at 20 K for the full distribution.

Detailed multiple shell analysis was conducted for the Tb-O distribution. Above ~180 K the Tb-O distribution is composed of multiple components which are not resolved. Below ~180 the distribution is resolvable into two Gaussian peaks. In Fig. 5, we show the full details of this bond distribution as a function of temperature in terms of the extracted distances, Debye Waller factors, and the ratio of the coordination number of the short bond relative to the long bond (short bond plus long bond coordination numbers constrained to sum to 8). The distances are well separated below ~180 K (with resolution limit constrained by the hard limit of the L2 edge above the L3 edge). The Debye-Waller factors maintain distinct values but approach each other at 10 K and below (last two points). With reduced temperature the ratios of the coordination numbers $N_{short}/N_{Long}$ approach each other rapidly and attain unity at ~65 K.

In order to visualize the qualitative trends, the Tb-O component of the radial distribution functions g(r) (in $\chi(k) = \int \chi_{th}(k,r)\, 4\pi r^2 g(r) dr$ given above) where examined . We display the evolution of the Tb-O radial distribution function from 470 K to 3 K in Fig. 6 Note that below ~180 K the total distribution is asymmetric but on crossing into the region where the Tb moments order magnetically (~10 K) the peaks become symmetric and sharpens.

In Fig. 7 we show the Pbam structure. The results derived from our XAFS analysis require a modification of this structure. A structure which preserves the local atomic order with respect to the Mn polyhedra and the second shell about Tb (Tb-Mn and higher shells) while hosting a distorted Tb-O distribution has limited possibilities. The data are consistent with rotations of the polyhedra about the c-axis generated by bucking at the "hinges" connecting them. The arrows in Fig. 7 illustrate possible bucking points which result in the motion of oxygen atoms towards one Tb site (and away from another) producing a split distribution. Other hinge points exist about any given c-axis chain of Tb ions. These rotations should be manifested by a lowering of the space group possibly in a manner compatible with the magnetic and ferroelectric state at low temperature and possibly in an increase in the cell size. Indeed, superlattice reflections have been observed at low temperature in the $DyMn_2O_5$ system revealing a doubling of the c-axis [20]. What should be noted is that unlike the case of polyhedra buckling in the simple $ABO_3$ perovskites [21], the asymmetry inherent in the c-axis ring of the polyhedra about the Tb sites does not preserve the Tb site symmetry on buckling. Buckling will allow a finite |**P**|.

Indeed in the $YMnO_3$ system, x-ray diffraction measurements and first principle LDA calculations reveal that the observed polarization (|**P**| =6 μC/cm$^2$) results from tilting of the of the Mn centered polyhedra and buckling of the Y-O planes with no significant off-center shift of the Mn ions [22] . The rotations of the MnOx polyhera suggested by the XAFS measurements in the $TbMn_2O_5$ systems will generate buckling of the "Tb-O planes" in an similar manner – putting this system in the same class as $YMnO_3$ of so called geometric ferroelectrics.

We note that in this system, finite |**P|** may not be solely due to Tb-O polarization but the Mn sites may contribute as proposed in previous work [11]. However, the Mn displacements are significantly smaller than those of the oxygen sites (due to rotations). The Tb site plays an important role which was not previously appreciated.

## IV. Summary

In summary, to first order the $MnO_6$ and $MnO_5$ polyhedra are found to be rigid, exhibiting a high average Einstein temperature. An anomalous temperature dependent Tb-O distribution is found. At high temperature it is broad but resolves into two distinct Tb-O peaks below ~180 K. The distributions sharpen below the Tb ion magnetic ordering temperature (~10 K). The distortions in the Tb-O distribution are consistent with rotations of the $MnO_x$ polyhedra about the c-axis and suggest that Tb-O bond polarization may play a significant role in the observed ferroelectric properties of this system. Re-O polarization may be important in the ferroelectric properties of the general $ReMn_2O_5$ system.

## Acknowledgment

This work was supported in part by NSF DMR-0512196 and NSF IMR Grant DMR-0083189. Work at Rutgers University was supported by NSF DMR-0520471. This work is dedicated to the memory of Ivy A. Charles-Tyson.

**Fig. 1**. XAFS data for the Mn K-Edge (thin line) and Tb L3-Edge (thick line) taken at 300K.

**Fig. 2**. Fourier transform of XAFS data taken at 300 K for the Mn K-Edge (top panel) and Tb L3-Edge (bottom panel) as thick lines. The dashed thin lines are fits to the data.

**Fig. 3**. Extracted $\sigma^2(T)$ for Mn-O fit to an Einstein model (dashed line) reveal a high Einstein ($\theta_E$) temperature indicating rigid $MnO_6$ and $MnO_5$ polyhedra to first order. The temperature axis is on a log scale.

**Fig. 4**. Local structure about the Tb sites. We compare (upper panel) data taken at 300 K (thick line) and at 20 K (thin line) with the crystallographic Pbam structure using reasonable global $\sigma^2$ (=0.0023 Å$^2$) for each atomic shell. For the 20 K data note that the first shell (Tb-O) does not follow the Pbam model. However, The Tb-Mn distribution (and higher shells) is well modeled and (lower panel) the temperature (log scale) dependence follows an Einstein model (dashed line).

**Fig. 5**. Details on the Tb-O distribution. The Tb-O distribution is split into two resolvable bonds below ~180 K (top panel) into long and short bond distribution. The corresponding Debye Waller factors are also shown in the middle panel. The ratio of the coordination numbers for the short and long bonds is plotted in the lower panel. Note that at 10 K and below the Debye Waller factors for both components attain the same value.

**Fig. 6**. Temperature dependent Tb-O distribution between 3 K and 470 K show in the top (460 to 200 K), middle (200 to 60K) and lower (60 to 3 K) panels. Below 180 K the Tb-O distribution is resolved into two well defined Tb-O distances which sharpen below 10 K.

**Fig. 7**. Crystal structure of $TbMn_2O_5$ based on Pbmn space group. Mn1 ($Mn^{4+}$) manganese sites occur at the center of the solid colored $MnO_6$ octahedra while Mn2 ($Mn^{3+}$) manganese site occur within the striped $MnO_5$ pyramids. The arrow indicate the "hinges" about which polyhedra may buckle producing the bimodal distribution in Tb-O distances.

**Fig. 1. Tyson *et al*.**

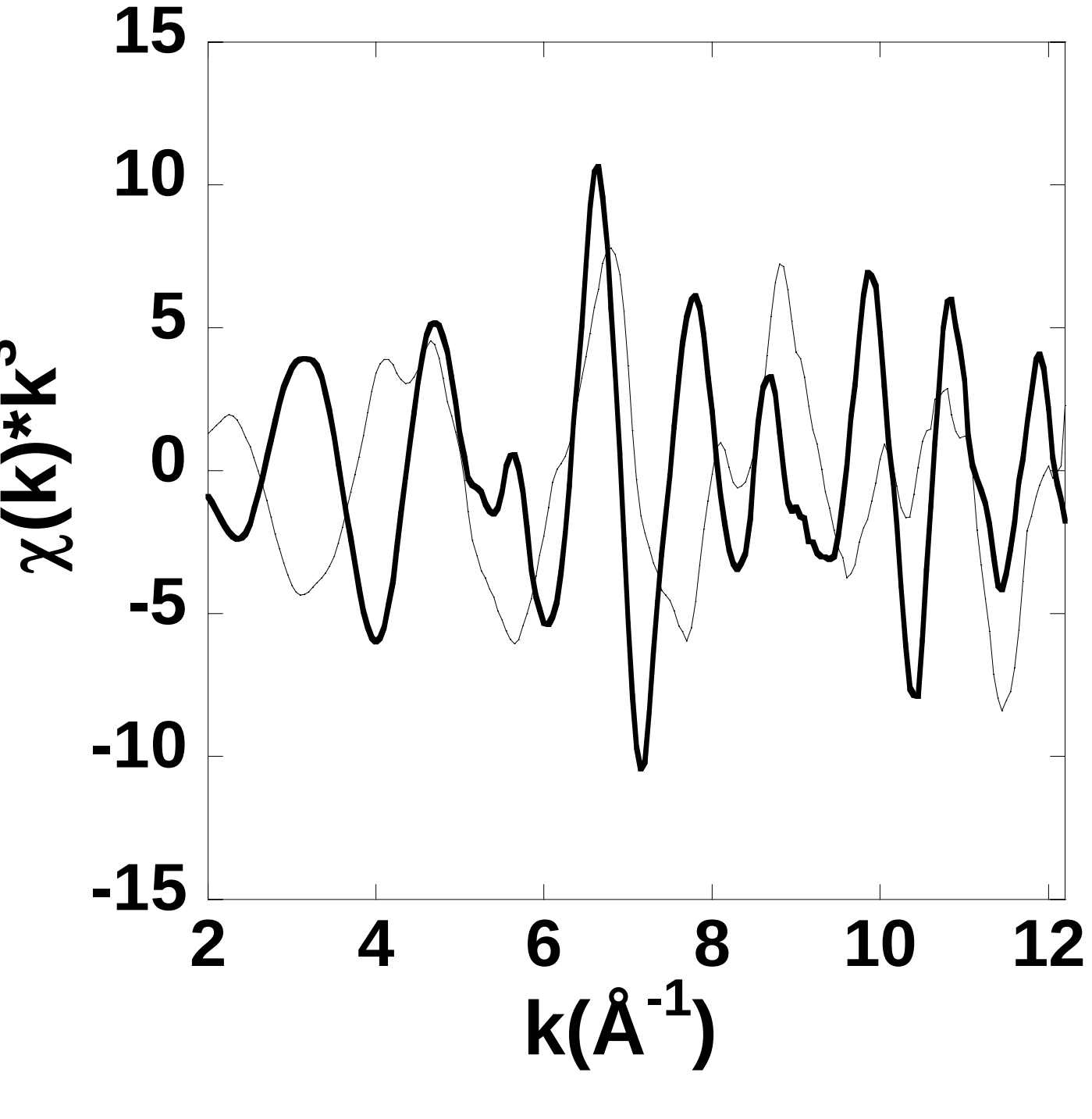

**Fig. 2. Tyson *et al*.**

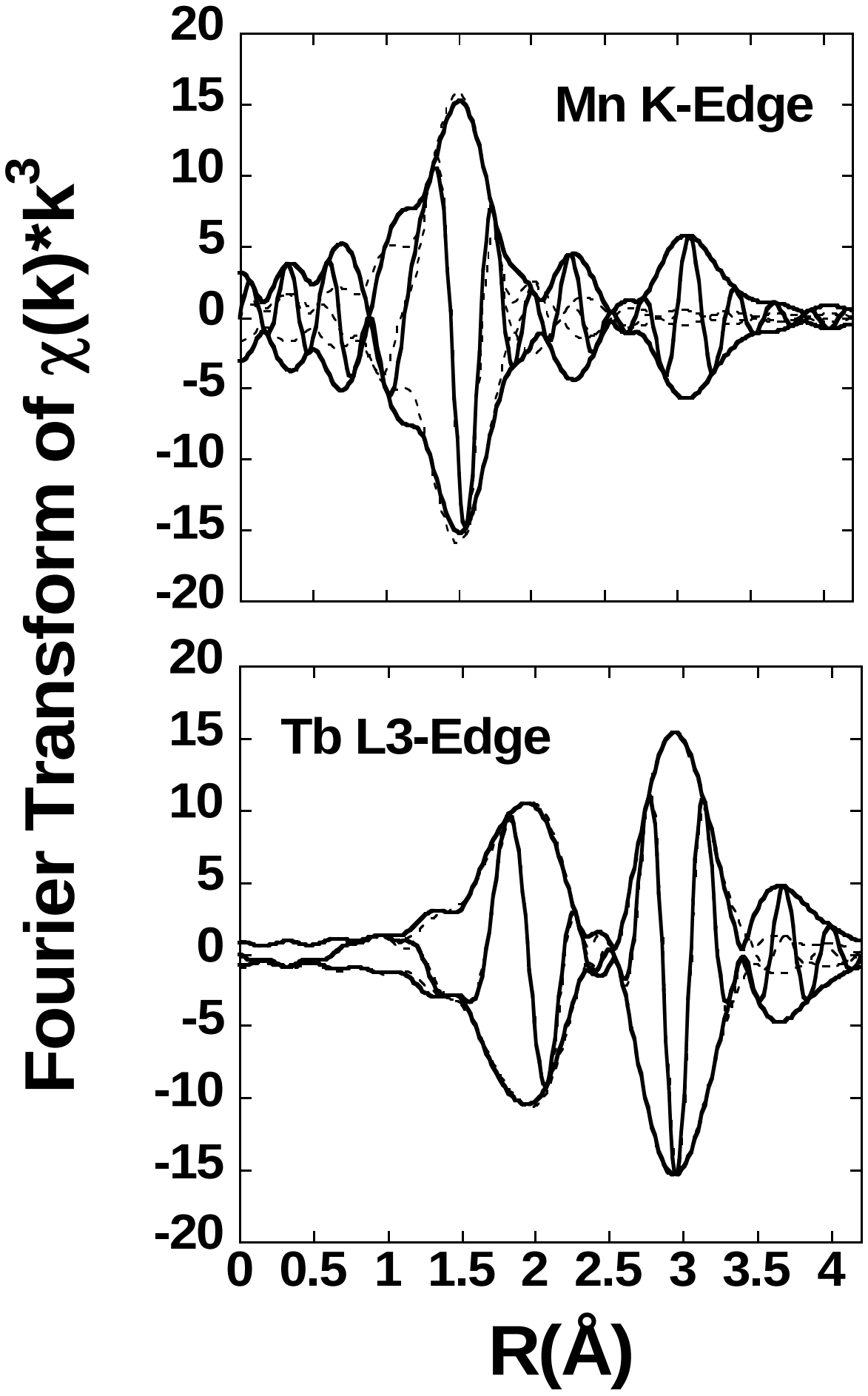

**Fig. 3. Tyson *et al*.**

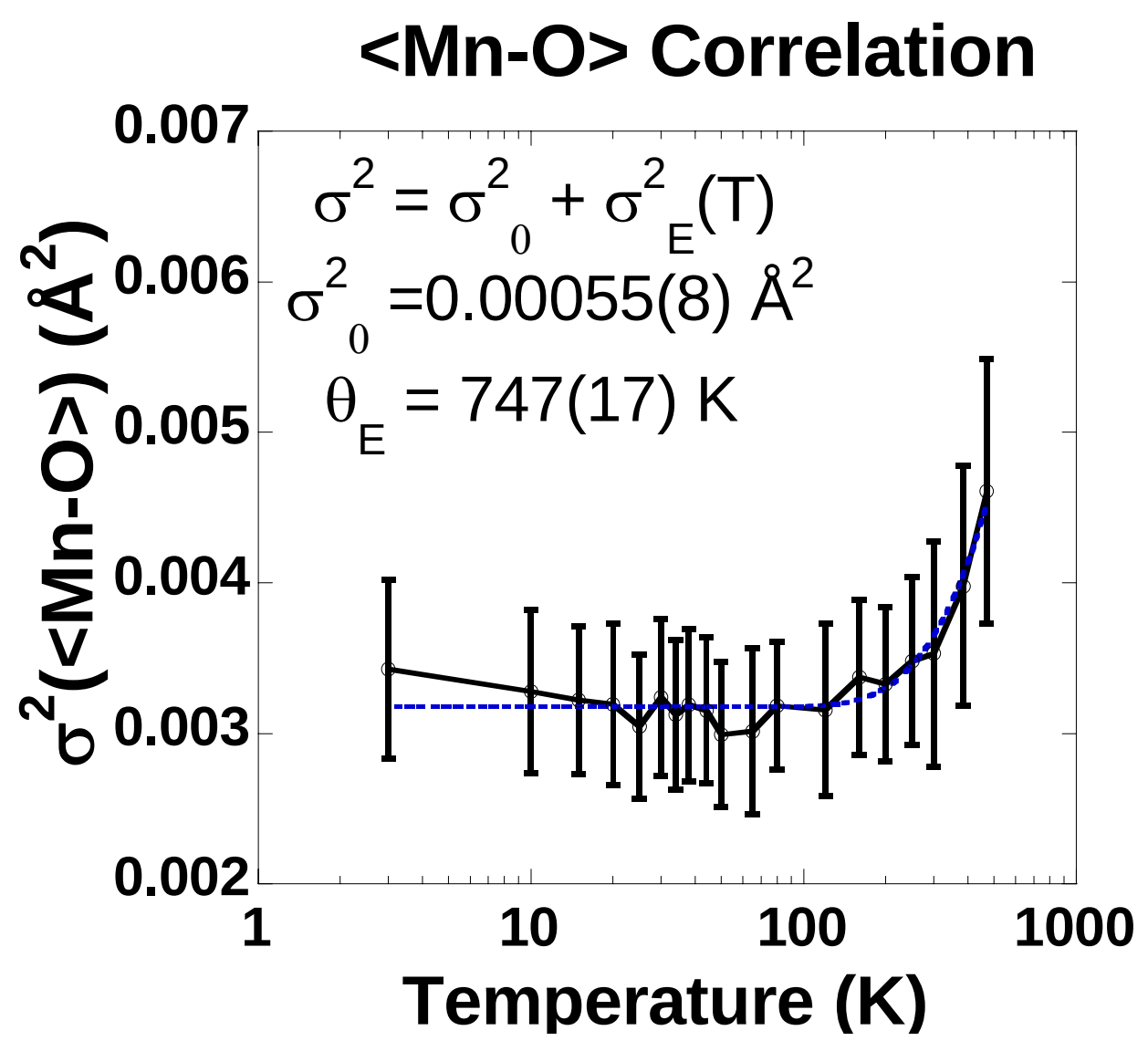

**Fig. 4**, **Tyson *et al*.**

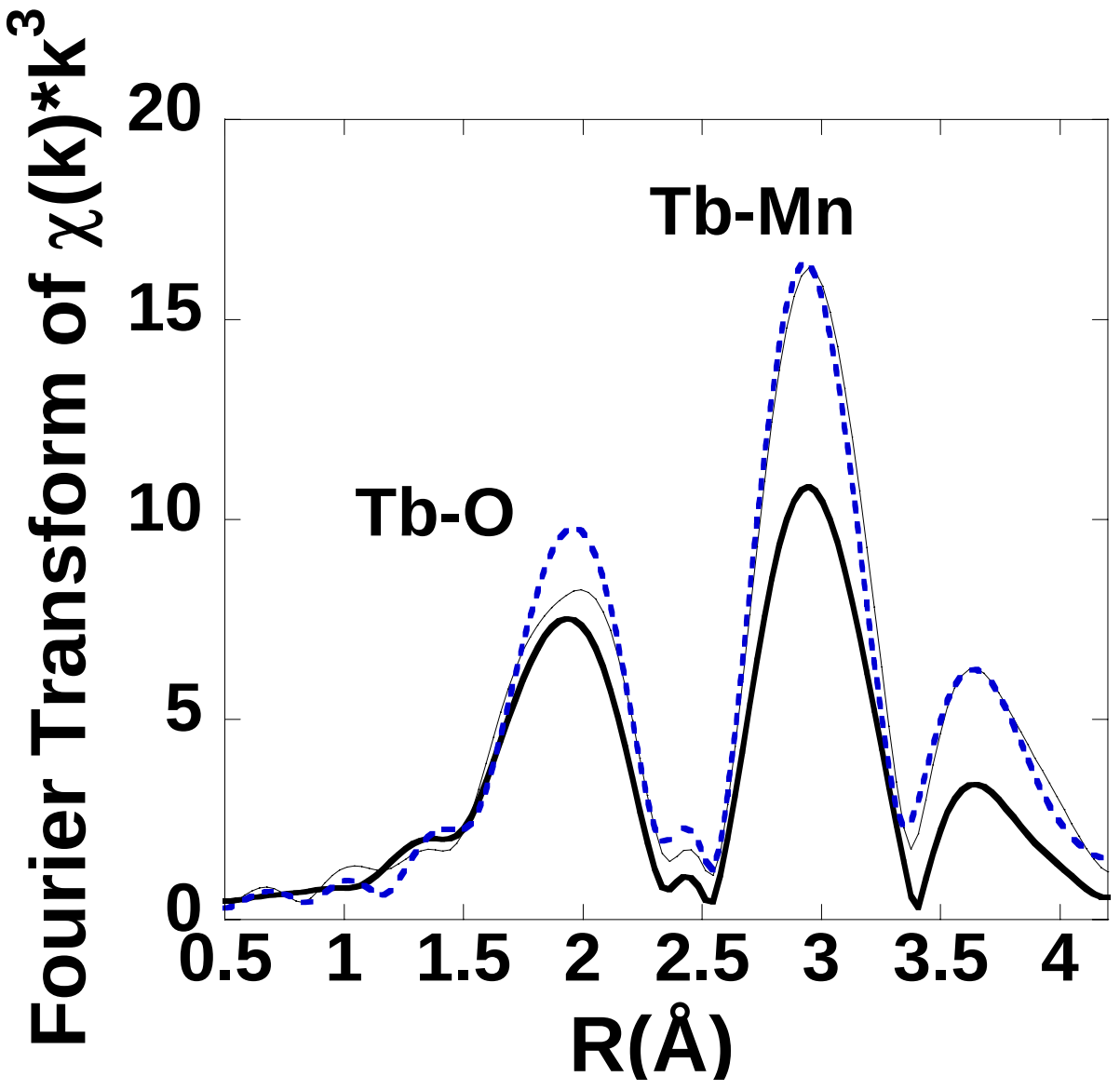

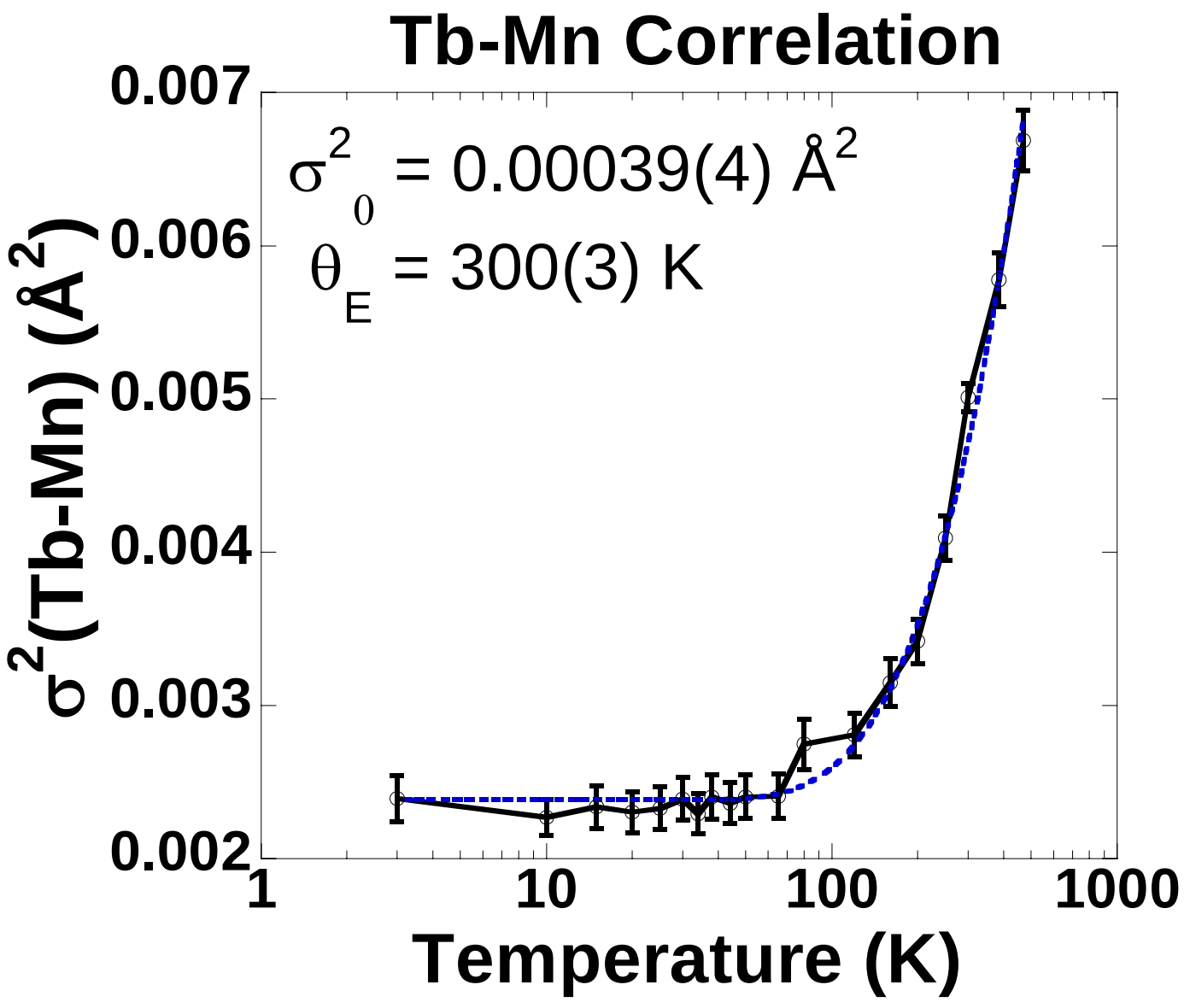

**Fig. 5**, **Tyson *et al*.**

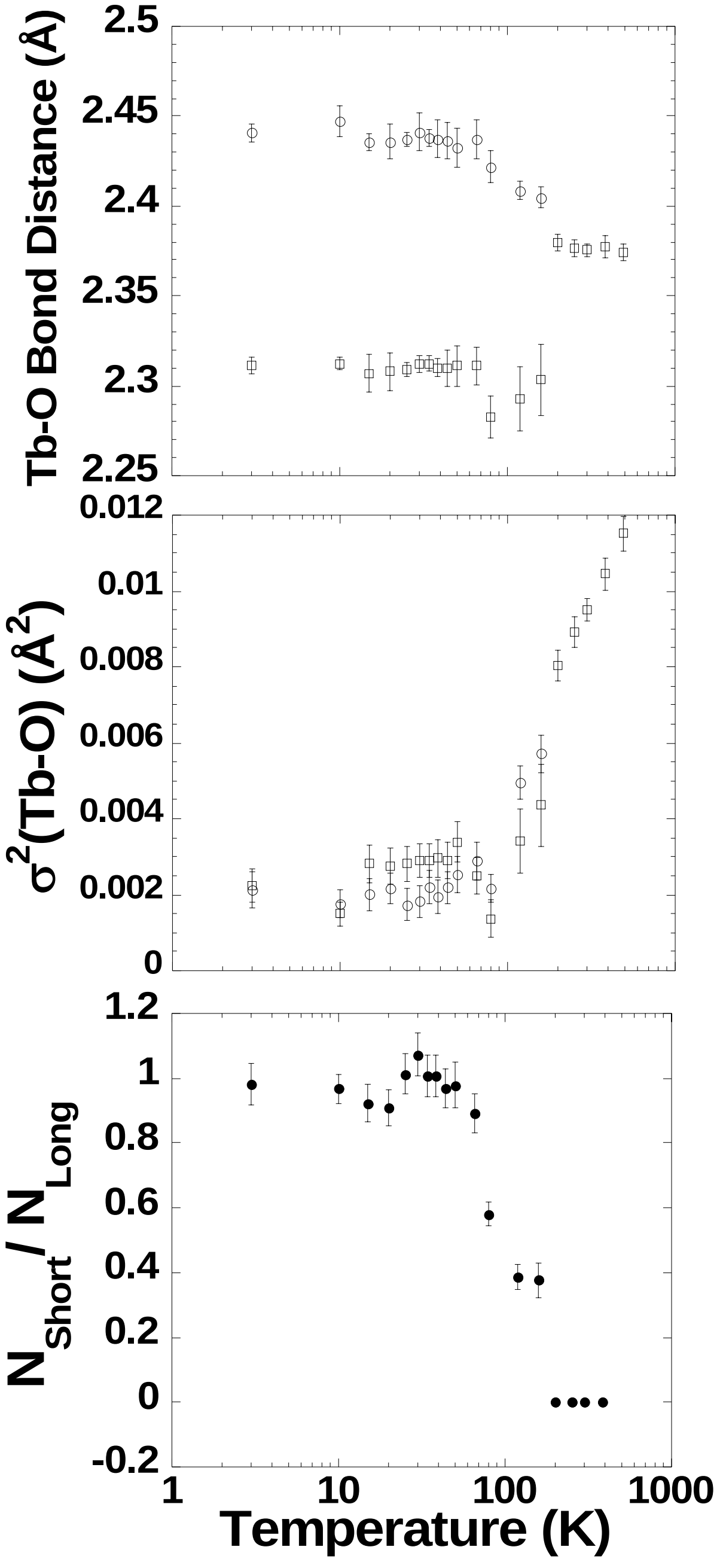

**Fig. 6, Tyson *et al*.**

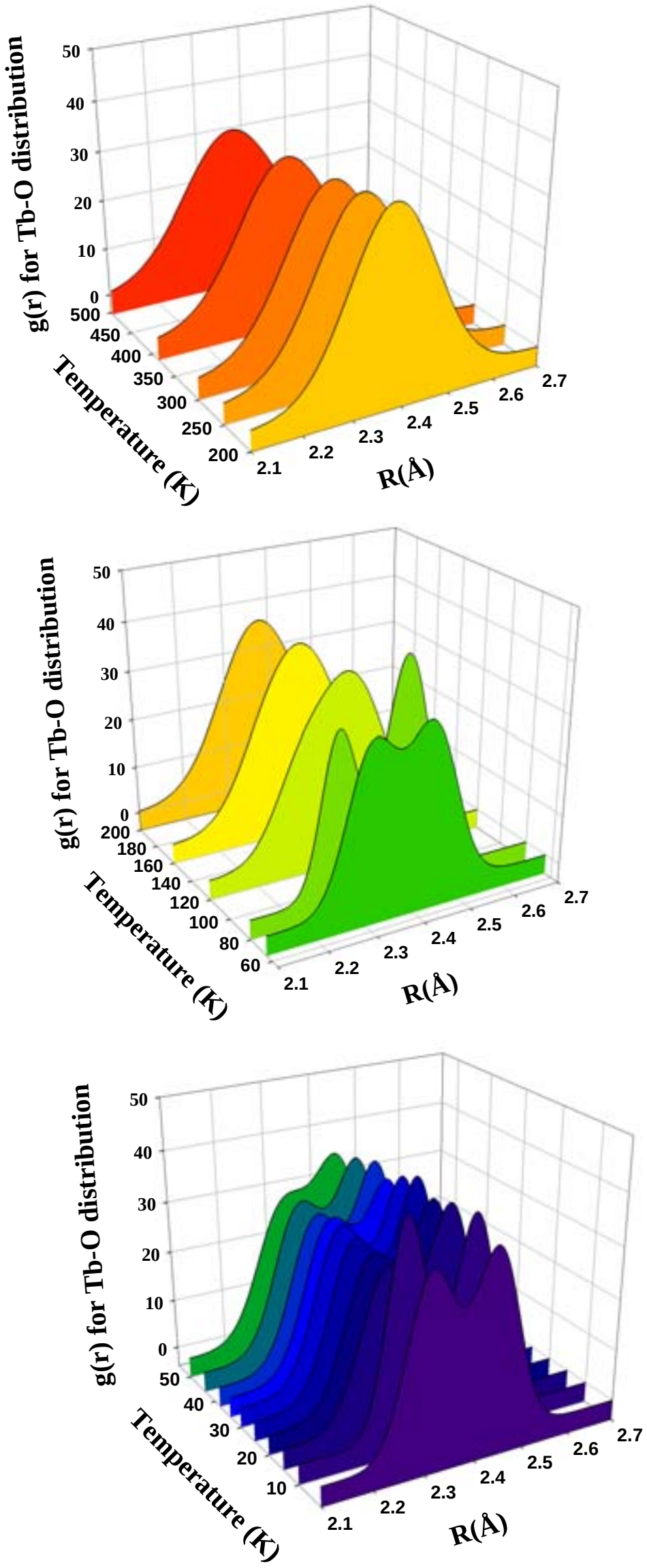

**Fig. 7. Tyson *et al*.**

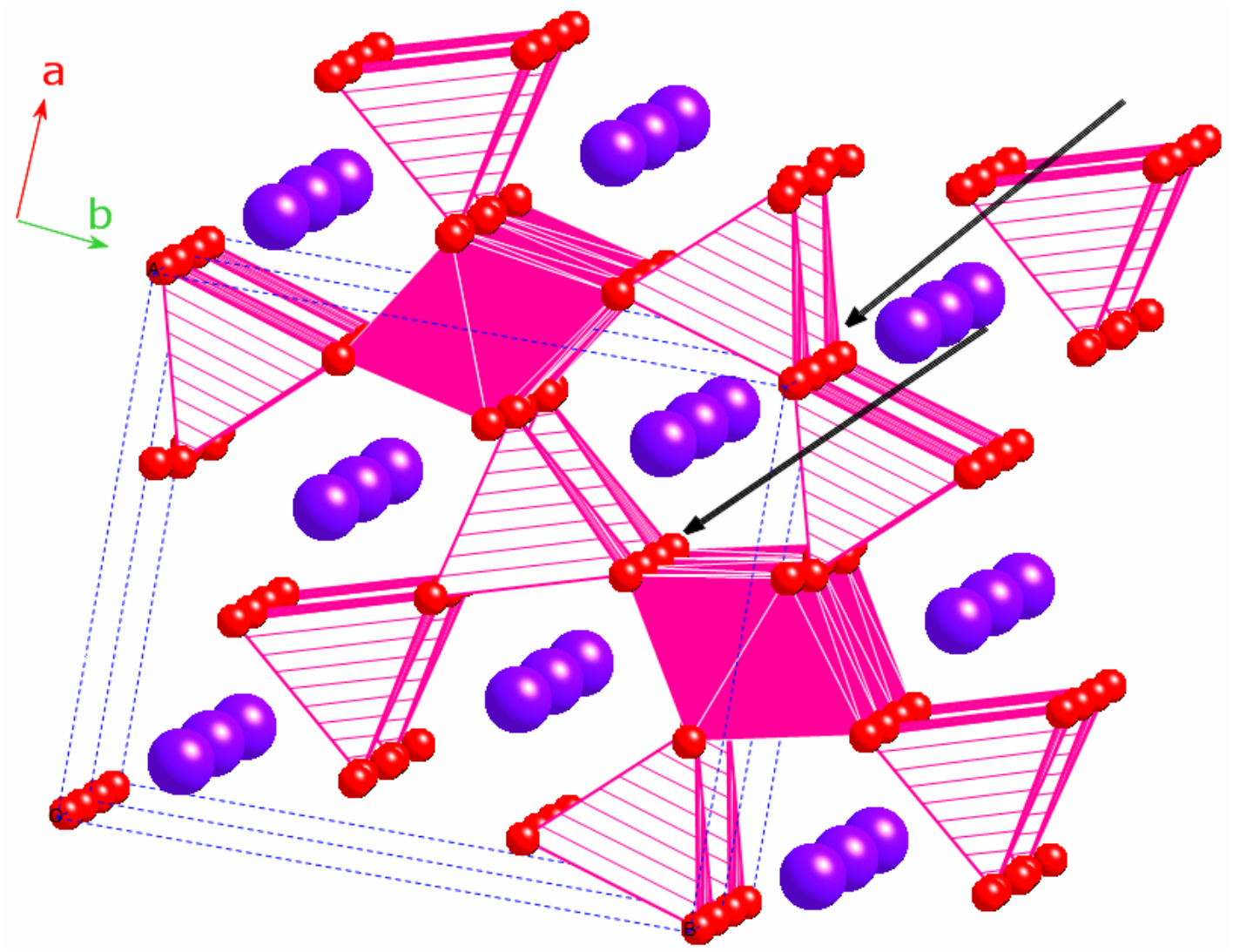